\documentclass[11pt]{article}
\usepackage{graphicx}
\newcommand{\odra}{O.~Dragoun}
\newcommand{\rys}{M.~Ry\v sav\'y}
\newcommand{\spa}{A.~\v Spalek}

\begin{document}
\newcommand{\beq}{\begin{equation}}
\newcommand{\eeq}{\end{equation}}
\newcommand{\bec}{\begin{center}}
\newcommand{\eec}{\end{center}}
\newcommand{\ben}{\begin{enumerate}}
\newcommand{\een}{\end{enumerate}}
\newcommand{\beqar}{\begin{eqnarray}}
\newcommand{\eeqar}{\end{eqnarray}}
\newcommand{\bit}{\begin{itemize}}
\newcommand{\eit}{\end{itemize}}
\newcommand{\Rb}{$^{83}$Rb}
\newcommand{\ikry}{$^{83m}$Kr}
\newcommand{\kry}{$^{83}$Kr}
\newcommand{\Ep}{E_{peak}}
\newcommand{\alf}{$\alpha$}
\newcommand{\bet}{$\beta$}
\newcommand{\gam}{$\gamma$}
\newcommand{\KAT}{KATRIN}
%
\begin{flushright} Report NPI ASCR \v Re\v z\\
TECH--05/2005\vspace{0.5ex}\\ \ \\
\end{flushright}
\bec \Large \bf
    Conversion electrons used to monitor the energy scale of electron
    spectrometer near tritium endpoint
      -- a simulation study
\eec
\bec
    \rys\footnote{e-mail: rysavy@ujf.cas.cz}\\
   \small Nuclear Physics Institute, Acad. Sci. Czech Rep., \\
   \small CZ--250 68 \v{R}e\v{z} near Prague, Czech Republic
\eec
\date{}
\begin{abstract}
Measurements of the endpoint region of the tritium \bet--decay
spectrum provides good possibility to determine neutrino mass.
This, however, needs a perfect monitoring of the spectrometer
energy scale. A parallel measurement of electron line of known
energy -- in particular the \ikry\ conversion K-line -- may serve
well to this purpose. The \Rb\ decaying to \ikry\ seems to be a
very suitable radioactive source due to its halflife of 86.2 day.
In this work, we determine the amount of \Rb\ which is necessary
for a successful monitoring.
\end{abstract}
%
\section{Introduction}   \label{s:intro}
In any spectroscopic experiment, good knowledge of spectrometer
energy scale is the {\em condition sine qua non} for a successful
performance. First, we must assure that the energy is really that
one which we think it is (as the instruments tell us it is); this
is called {\em calibration}. Second, this condition must be
fulfilled during the whole duration of the measurement; this is
the aim of the energy {\em monitoring}.

One of the most reliable ways of monitoring is to measure some
spectrum line (preferably a well separated one) of a well known
energy. When measuring the tritium spectrum by a ``main''
spectrometer it is reasonable to use a smaller monitoring
spectrometer -- measuring the K-conversion line of \ikry\ --
connected to the same controls as the ``main'' one.
 As the source,
\Rb\ decaying to \ikry\ may be utilized. (Note that the idea to
calibrate and/or monitor the tritium spectrometer using \ikry\ is
nothing new -- see e.g. \cite{Rob91} and references cited
therein.) Another possibility might be to measure photoelectrons
from Co initiated by $^{241}$Am 26~keV \gam-rays.

As stated before, we need the stable energy scale. In other words,
if any change appeared in it, we need to recognize it immediately.
In this work we determine how much krypton activity (and then, how
much \Rb\ activity, since the krypton and rubidium decays are in
balance with each other) is necessary to fulfill such a task in a
real situation. We use the data which are expected for the
prepared experiment \KAT\ and are published in \cite{LoI,DesRep}.
 We simulate
\ikry\ conversion lines with intentionally wrong (mutually
shifted) energies and then apply statistical tests to find the
minimum shift which makes the spectra distinguishable. Repeating
this for various activities and/or exposure times we get relations
which enable one to optimize the monitoring.
\section{Method}
The shape of the conversion line is described by a Lorentzian
curve
\beq
    K(E) = \frac{1}{2\pi}\times
    \frac{\Gamma}{(E-\Ep)^2 + (\Gamma/2)^2}\ ,
\eeq
where $\Ep$ and $\Gamma$ are the line energy and FWHM,
respectively.

This line shape is measured by an integrating spectrometer, the
response function of which is \cite{DesRep}
\beq
   RpF(E,qU) = \left\{
   \begin{array}{ll} 0 &\ \ 0\le E\le qU\\
   \frac{1-\sqrt{1-\frac{E-qU}{E}\times \frac{B_S}{B_A}}}{1-
       \sqrt{1-\frac{\Delta E}{E}\times \frac{B_S}{B_A}}}  & \ \ qU\le
       E\le qU+\Delta E\\
   1 & \ \ qU+\Delta E\le E
   \end{array}
   \right.
\label{e:RpF}
\eeq
where $B_S, B_A$ are the spectrometer magnetic field settings at
the source and in the analyzing plane, and $\Delta E$ is the
spectrometer resolution. The registered spectrum is then given by
summing
\beq
   S(E) = \int_0^\infty RpF(T,E)\: K(T)\: {\rm d}T.
\eeq
This integral may be partially simplified
\beq
  S(E) = \int_E^{E+\Delta E} RpF(T,E)\: K(T)\: {\rm d}T +
  \frac{1}{2} - \frac{1}{\pi} \arctan\frac{2(E+\Delta
  E-\Ep)}{\Gamma}.
\eeq
The first part must be calculated numerically. However, the middle
part of the response function, Eq.(\ref{e:RpF}), may be well
approximated by a straight line. For such a case we get an
analytical formula
\beqar
   \int_E^{E+\Delta E} RpF(T,E)\: K(T)\: {\rm d}T & = & \frac{1}{2\pi
   \Delta E}\times \left\{\frac{\Gamma}{2}\ln\frac{(E-\Ep+\Delta
   E)^2 + (\Gamma/2)^2}{(E-\Ep)^2 + (\Gamma/2)^2} \right. \nonumber \\
  & & \left. - 2(E-\Ep)\left[\arctan\frac{2(E-\Ep+\Delta E)}{\Gamma} \right. \right. \label{e:anal}
\\
  & & \left. \left. -
  \arctan\frac{2(E-\Ep)}{\Gamma}\right]\right\}.   \nonumber
\eeqar
Using both the numerical integration and this formula, we can
simply check the precision of our numerical procedures. We did so
for the values of the \ikry\ conversion K-line ($\Ep$=17823.8~eV,
$\Gamma$=2.8~eV) \cite{Pic92}, the spectrometer resolution of
$\Delta E$=2~eV, and set of 5 values of $E$ (17810 to 17830~eV
with the step of 5~eV). Using the simplest possible integration
method -- summation with the step of 0.01~eV -- the results agreed
with the analytical ones of Eq.(\ref{e:anal}) up to more than 5
digits.
\section{Calculations}
To get reasonable estimates of the sensitivity we performed a lot
of simulations of the \ikry\ K-line spectra. Moreover, the
simulations were done assuming similar conditions under which such
a spectrum was really measured \cite{Kas05} in June 2005 in Mainz.
This assures that the results might be directly applicable in
practice.

In our simulations we did not cover the whole measured energy
interval \cite{Kas05} but the most sensitive part only, usually
between 17819 and 17825~eV.  Since the linewidth fitted in
\cite{Kas05} was $\Gamma_{fit}$ = 3.08~eV, we repeated the
simulation for both this value and the value $\Gamma$ = 2.8~eV of
\cite{Pic92}. (In all cases, the realistic magnetic fields setting
of  $B_S/B_A$$\approx$8000 was used.) The difference between the
two sets of results turned out to be negligible. In the following,
then, we present the results obtained with $\Gamma$= 2.8~eV only.

\begin{figure}[h]
  \centerline{\includegraphics[clip=on,angle=0,width=10cm]{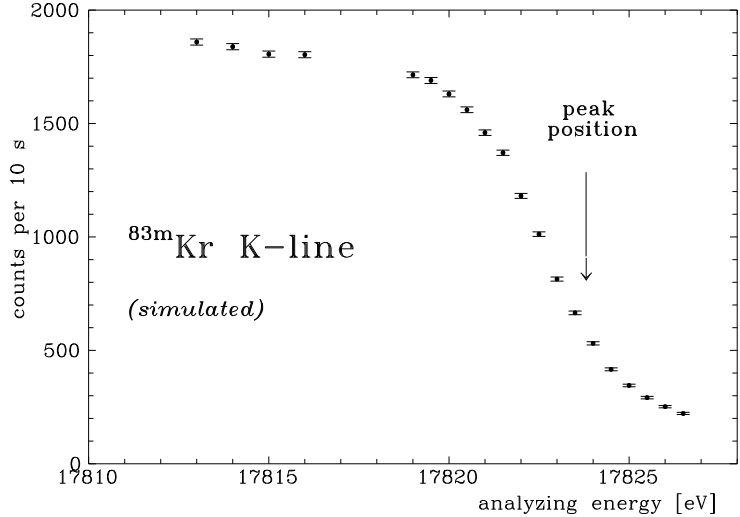}}
\caption{An example of simulated K-line of \ikry. The \ikry\
activity of 190~kBq and exposure of 10 s per point is assumed.
Analyzing energy is the quantity $qU$ of Eq.(\ref{e:RpF}). The
reason to include the four leftmost points is given later in the
text.} \label{f:simspe}
\end{figure}

In one run, always ten spectra were simulated having identical
amplitudes and linewidths but mutually shifted by a fixed energy
interval $\delta E$. An example of such simulated spectrum is at
the Fig.\ref{f:simspe}.
Then the statistical tests \cite{Dra97} were applied to the
spectra set. (We remind the reader here that those tests compare
every spectrum with each other to verify whether they were
measured {\em under the same conditions}.) We seek the smallest
$\delta E$ for which the tests are able to distinguish two
neighbouring spectra. Thus we repeat the above procedure with
decreasing $\delta E$ until the neighbouring spectra cease to be
distinguishable.

We accept the spectra as distinguishable if -- in the particular
run -- most of the corresponding \cite{Dra97} $\chi^2$-values (per
one d.o.f.) are greater than 1.5 and/or most of the corresponding
\cite{Dra97} Student-test-values are in absolute value greater
than 1. (The rather high value of $\chi^2$=1.5 was chosen due to
the following reason: having $\chi^2 \ge 1.5$, we can be {\em
almost sure} that the spectra compared are \underline{really}
different, i.e. the difference is not due to statistics.)

Here is one thing which must be kept in mind. The tests
\cite{Dra97} check only ``measurements {\em were} or {\em were
not} done under the same conditions''. So a change of detection
efficiency would also lead to the `not same conditions' signal
thus imitating a spectrum shift. Fortunately, the realization of
the tests (the software) contains a branch performing a
normalization of the spectra being compared to each other. If the
`not same conditions' signal is caused by a detection efficiency
change, this normalization releases it while if the cause is an
energy scale change, it does not. This allows us to discern the
false and true effects. And this is also the reason why we
included several low-energy points (four leftmost ones in Fig.
\ref{f:simspe}) into simulations. They are out of the region
sensitive to energy shift but they are sensitive to amplitude
change which strongly helps to the normalization procedure.
\section{Results}
First, we simulated the situation corresponding to the measured
spectrum \cite{Kas05}, i.e. the \ikry\ activity $A$=1.9~kBq and
exposure of 350~s per point. The 28 points in the most sensitive
region (in particular from 17819 to 17824.75~eV with the step of
0.25~eV) were used.

Under these conditions, we are able to distinguish the spectra
which were shifted by about 100~meV or more which is not too
satisfactory. Fortunately, there are possibilities to improve it.

\begin{figure}[h]
  \centerline{\includegraphics[clip=on,angle=0,width=10cm]{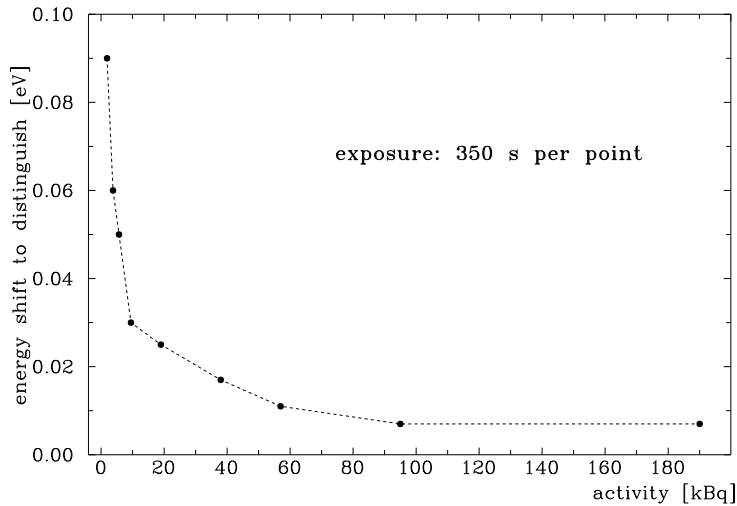}}
\caption{Dependence of the minimum detectable spectra shift on the
\ikry\ activity.} \label{f:actdep}
\end{figure}

First, we were increasing the activity up to hundredfold larger
than the original one. As expected, smaller and smaller shifts
could be resolved. The results are given in Fig. \ref{f:actdep}.
The recognizable shift of less than 10~meV is excellent. However,
the overall measurement time of about $2\frac{3}{4}$ hour is
rather long for practical monitoring.

\begin{figure}[h]
  \centerline{\includegraphics[clip=on,angle=0,width=10cm]{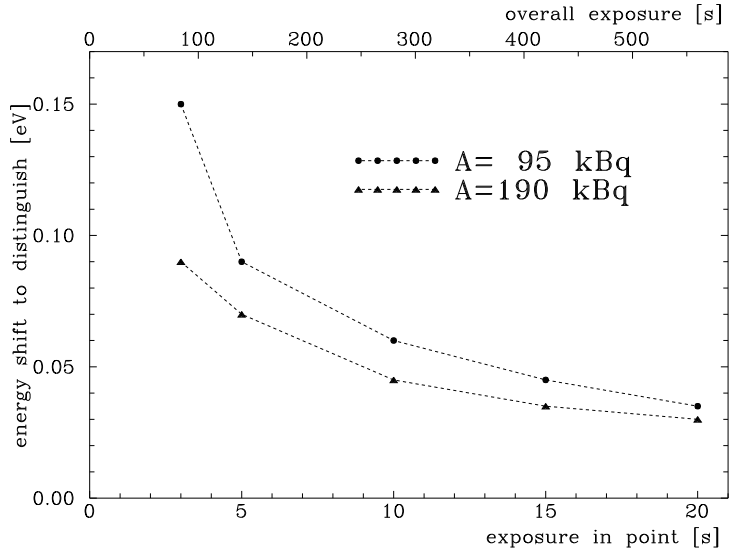}}
\caption{Dependence of minimum detectable spectra shift on
exposure for two \ikry\ activities.} \label{f:timdep}
\end{figure}

As the next step we performed the simulations for various
(shorter) exposures. As for the activities, we used 95 and
190~kBq; this corresponded to 50$\times$ and 100$\times$,
respectively, the activity \cite{Kas05} and is realizable without
difficulties. The results are in Fig. \ref{f:timdep}. It turns out
that quite satisfying sensitivity to shift of about 30~meV is
reached after the exposure of some 10 minutes. A longer exposure
brings very little gain.

Finally, we tried to reduce the number of measured points in a
hope to further reduce the necessary measurement time. We took
only 15 points instead of 28 in the same region as before (i.e. we
applied the step of 0.5~eV) and repeated all the previous
simulations. The dependence on activity (Fig.~\ref{f:actdep})
practically did not change. For the exposure dependence, however,
approximately twofold exposure per point was needed to achieve the
same sensitivity as before. Then the overall time needed to
measure the spectrum remains the same.
\section{Conclusions}
The measurement of \ikry\ conversion K32--line is a very good
method to monitor the energy scale of electron spectrometer near
the tritium endpoint. The necessary activity of \Rb\ is reasonable
and there are no difficulties to produce it. However, for
practical utilization there still remains the question of
long--term stability of the \Rb/\ikry\ composition
\cite{Kas05,Ven05} (i.e. of spontaneous releasing of the krypton
into vacuum) to be solved.
%
\section*{Acknowledgement}
This work was partly supported by the Grant Agency of the Czech
   Republic under contracts No. 202/02/0889, 202/06/0002 and by the ASCR project AV0Z10480505
%

%

\begin{thebibliography}{99}
%
\bibitem{Rob91} R.G.H. Robertson, T.J. Bowles, G.J. Stephenson,
    Jr., D.L. Wark, J.F. Wilkerson: Phys. Rev. Lett. {\bf 67} (1991)
    957.
\bibitem{LoI} A.~Osipowicz et al. (\KAT\ Collaboration),
     {\tt arXiv:hep-ex/0109033} (2001).
\bibitem{DesRep} \KAT\ Collaboration, ``\KAT\ Design Report'',
    Report FZKA 7090, NPI \v Re\v z EXP-01/2005, MS-KP-0501.\\
    {\tt http://www-ik.fzk.de/katrin/publications/documents/FZKA7090.pdf
    http://bibliothek.fzk.de/zb/berichte/FZKA7090.pdf}.
\bibitem{Pic92} A. Picard et al.: Z. Physik {\bf A342} (1992) 71.
\bibitem{Kas05} J. Ka\v spar: private communication.
\bibitem{Dra97} \odra, \rys, \spa: Nucl. Instr. Meth. {\bf A391}
    (1997) 345.
\bibitem{Ven05} D. V\'enos, \spa, O. Lebeda, M. Fi\v{s}er: Appl.
Radiat. Isot. {\bf 63} (2005) 323.
\end{thebibliography}
\end{document}